# Article information

## Article title
Datasets on materials research of hard ferromagnet in TM-Fe-Si (TM=Ti, Zr, Hf, V, Nb, and Ta) ternary systems

## Authors
Ryusei Ishibashi, Jiro Kitagawa*

## Affiliations
Department of Electrical Engineering, Faculty of Engineering, Fukuoka Institute of Technology, 3-30-1 Wajiro-higashi, Higashi-ku, Fukuoka 811-0295, Japan

## Corresponding author's email address and Twitter handle
j-kitagawa@fit.ac.jp

## Keywords
Permanent magnet
X-ray diffraction pattern
Microstructure
Magnetization
Coercive field

## Abstract
The datasets presented in this article are related to materials research on hard ferromagnet in TM-Fe-Si (TM=Ti, Zr, Hf, V, Nb, and Ta) ternary systems. The motivation for data collection is based on the research paper entitled "Novel hard magnetic phase with $Zr_{11.5}Fe_{53}Si_{35.5}$ composition". The datasets are composed of scanning electron microscope images, X-ray diffraction (XRD) patterns, and magnetization data for $TM_7Fe_{52}Si_{41}$ annealed at 1050 ℃. The chemical compositions of constituent phases were determined by an energy dispersive X-ray spectrometer (EDS). The phase analysis was performed using XRD and EDS results. The Curie temperature of each sample was obtained using magnetization data, and the coercive field was determined for hard ferromagnet samples $Zr_7Fe_{52}Si_{41}$ and $Hf_7Fe_{52}Si_{41}$. The datasets would be useful for developing an Fe-based rare-earth-free permanent magnet, which is one of the central issues of materials science.

## Specifications table

| | |
|---|---|
| **Subject** | Metals and alloys |
| **Specific subject area** | Phase analyses and ferromagnetic properties of Fe-based compounds |

| **Type of data** | Image<br>Figure |
|---|---|
| **How the data were acquired** | Scanning electron microscope (SEM), energy dispersive X-ray spectrometer (EDS), powder X-ray diffractometer (XRD), vibrating sample magnetometer (VSM) |
| **Data format** | Raw<br>Analyzed |
| **Description of data collection** | Each sample was prepared by arc melting constituent elements and annealed at 1050 ℃ for one day in an evacuated quartz tube. SEM and EDS results were obtained for samples with a polished surface. XRD patterns were taken for powdered samples. VSM data were collected using bulk samples between 50 K and 800 K. |
| **Data source location** | · Institution: Fukuoka Institute of Technology<br>· City/Town/Region: Fukuoka<br>· Country: Japan |
| **Data accessibility** | Data are with the article. The raw data are in the Mendeley Data repository.<br><br>https://data.mendeley.com/datasets/gp8rkw2k6v/2 |

**Value of the data**

· The datasets are useful for elucidating the phase relation in TM-Fe-Si (TM=Ti, Zr, Hf, V, Nb, and Ta) ternary diagrams and designing magnetic properties of an Fe-based compound.
· The data would help the researchers working on the field of rare-earth-free magnetic materials to develop a novel permanent magnet.
· The data can be used for simulating the phase diagram of TM-Fe-Si ternary system and creating a strategy of coercive field enhancement of rare-earth-free magnets.

**Objective**

Rare-earth-free magnetic materials like Fe-based compounds are good candidates for low-cost magnets. It is beneficial to collect phase analyses and fundamental magnetic property data obtained through materials research on new Fe-based compounds. The datasets of structural characterization and magnetization in Fe-based materials are indispensable for developing a new rare-earth-free permanent magnet.

**Data description**

Recently Yamamoto et al. have discovered a novel hard magnetic phase with $Zr_{11.5}Fe_{53}Si_{35.5}$ composition [1]. The samples are multiphase, and one of the starting compositions showing superior hard magnetic properties is $Zr_7Fe_{52}Si_{41}$. The microstructure analysis suggests that $Zr_7Fe_{52}Si_{41}$ contains $Zr_{11.5}Fe_{53}Si_{35.5}$, FeSi, and $Fe_5Si_3$. Motivated by the paper [1], we collected datasets of SEM images, EDS results, XRD patterns, and magnetization to investigate the TM (TM=Ti, Zr, Hf, V, Nb, and Ta) dependences of phase relation and magnetic properties in $TM_7Fe_{52}Si_{41}$. Figs. 1(a) to 1(f) show the SEM images of $TM_7Fe_{52}Si_{41}$ (TM=Ti, Zr, Hf, V, Nb, and Ta), respectively. Each sample is composed of two or three phases, and the detected chemical compositions are noted in the figure. All samples contain the FeSi phase. In TM=Ti, Zr, and Hf samples, the phases close to the composition of $Zr_{11.5}Fe_{53}Si_{35.5}$ are detected ($Ti_{10.0}Fe_{52.5}Si_{37.5}$, $Zr_{10.7}Fe_{56.4}Si_{33.0}$, $Hf_{9.8}Fe_{54.9}Si_{35.4}$). The $Hf_7Fe_{52}Si_{41}$ sample exhibits a small amount of $Hf_{15}Fe_{44}Si_{41}$ with an unknown structure. For TM=V, Nb, and Ta samples, $Fe_5Si_3$ phase with the $Mn_5Si_3$-type structure is observed ($V_{9.9}Fe_{52}Si_{38}$, $Nb_{3.0}Fe_{59.6}Si_{37.3}$, and $Ta_{1.7}Fe_{61}Si_{37}$). $Fe_5Si_3$ might be partially substituted by the TM atom. $Nb_7Fe_{52}Si_{41}$ and $Ta_7Fe_{52}Si_{41}$ samples possess NbFeSi with the TiNiSi-type structure and the C14 Laves phase $Ta_{24}Fe_{41}Si_{34}$, respectively.

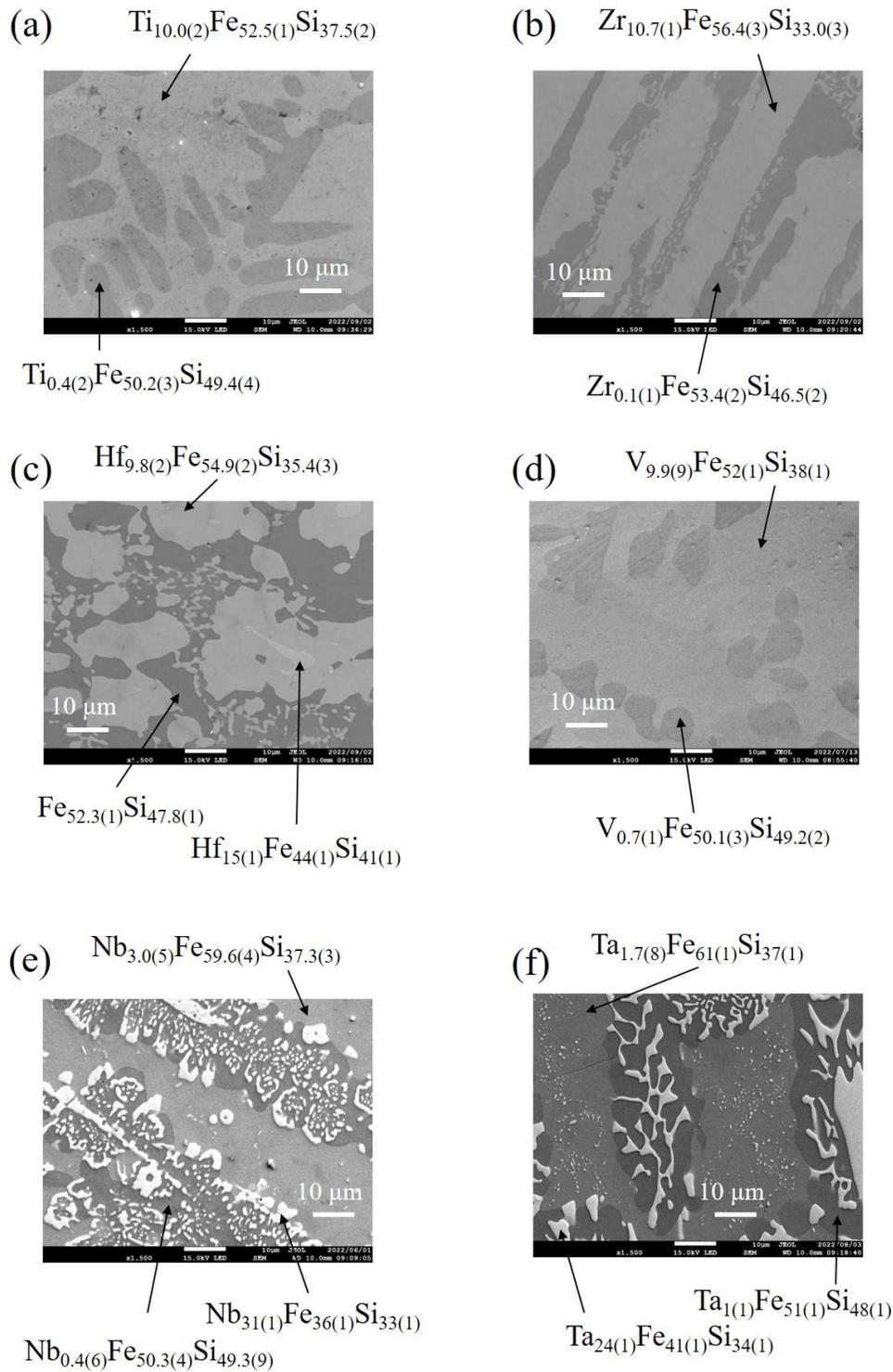

Fig.1 SEM images of (a) Ti$_7$Fe$_{52}$Si$_{41}$, (b) Zr$_7$Fe$_{52}$Si$_{41}$, (c) Hf$_7$Fe$_{52}$Si$_{41}$, (d) V$_7$Fe$_{52}$Si$_{41}$, (e) Nb$_7$Fe$_{52}$Si$_{41}$, and (f) Ta$_7$Fe$_{52}$Si$_{41}$, respectively. In each image, the chemical compositions of phases determined by EDS are also noted.

The XRD patterns of TM$_7$Fe$_{52}$Si$_{41}$ (TM=Ti, Zr, Hf, V, Nb, and Ta) are presented in Figs. 2(a) to 2(f). The simulation patterns of some compounds with known crystal structures in the ICSD database are added. The coll code of the ICSD database for each simulation pattern is given in the figure. The peak assignments are carried out taking into account the EDS results (see also Figs. 1(a) to 1(f)). All samples contain the XRD pattern of FeSi. In the Ti$_7$Fe$_{52}$Si$_{41}$ sample, the diffraction peaks other than the FeSi XRD pattern would originate from Ti$_{10.0}$Fe$_{52.5}$Si$_{37.5}$ with an unknown structure (▼ in Fig. 2(a)). Although the crystal structure of Zr$_{11.5}$Fe$_{53}$Si$_{35.5}$ composition is unclear, the paper [1] has reported the XRD pattern of the phase. We also confirmed the XRD pattern of Zr$_{11.5}$Fe$_{53}$Si$_{35.5}$ phase (▼ in Fig. 2(b)). Almost the same pattern is detected in the Hf$_7$Fe$_{52}$Si$_{41}$ sample, as shown by ▼ in Fig. 2(c). The diffraction peaks other than FeSi and Hf$_{9.8}$Fe$_{54.9}$Si$_{35.4}$ XRD patterns would be assigned as the Hf$_{15}$Fe$_{44}$Si$_{41}$ phase. We note that the intensity of the reflections at 35-38° attributed to the Zr$_{11.5}$Fe$_{53}$Si$_{35.5}$ phase in Hf$_7$Fe$_{52}$Si$_{41}$ is larger compared to Zr$_7$Fe$_{52}$Si$_{41}$ (see ▼ in Figs. 2(b) and 2(c)). The unknown Hf$_{15}$Fe$_{44}$Si$_{41}$ phase may contribute to 35-38° reflections marked by ▼. While the chemical composition of Ti$_{10.0}$Fe$_{52.5}$Si$_{37.5}$ is similar to that of Zr$_{10.7}$Fe$_{56.4}$Si$_{33.0}$ or Hf$_{9.8}$Fe$_{54.9}$Si$_{35.4}$, the XRD pattern of Ti$_{10.0}$Fe$_{52.5}$Si$_{37.5}$ is different from that of Zr$_{10.7}$Fe$_{56.4}$Si$_{33.0}$ or Hf$_{9.8}$Fe$_{54.9}$Si$_{35.4}$. The diffraction peaks of V$_7$Fe$_{52}$Si$_{41}$ shown in Fig. 2(d) can be assigned as Fe$_5$Si$_3$ or FeSi phase. In Nb$_7$Fe$_{52}$Si$_{41}$, the diffraction peaks due to NbFeSi with the TiNiSi-type structure appear in addition to the XRD patterns of Fe$_5$Si$_3$ and FeSi (Fig. 2(e)). As indicated in Fig. 2(f), the XRD pattern of Ta$_7$Fe$_{52}$Si$_{41}$ can be explained by the superposition of those of Ta$_{33}$Fe$_{45}$Si$_{22}$ (C14 Laves phase), Fe$_5$Si$_3$, and FeSi. The chemical composition for the C14 Laves Ta$_{33}$Fe$_{45}$Si$_{22}$ in the simulation patterns is near to Ta$_{24}$Fe$_{41}$Si$_{34}$ detected by EDS (see also Fig. 1(f)). The raw data of experimental XRD patterns are stored in the Mendeley Data repository.

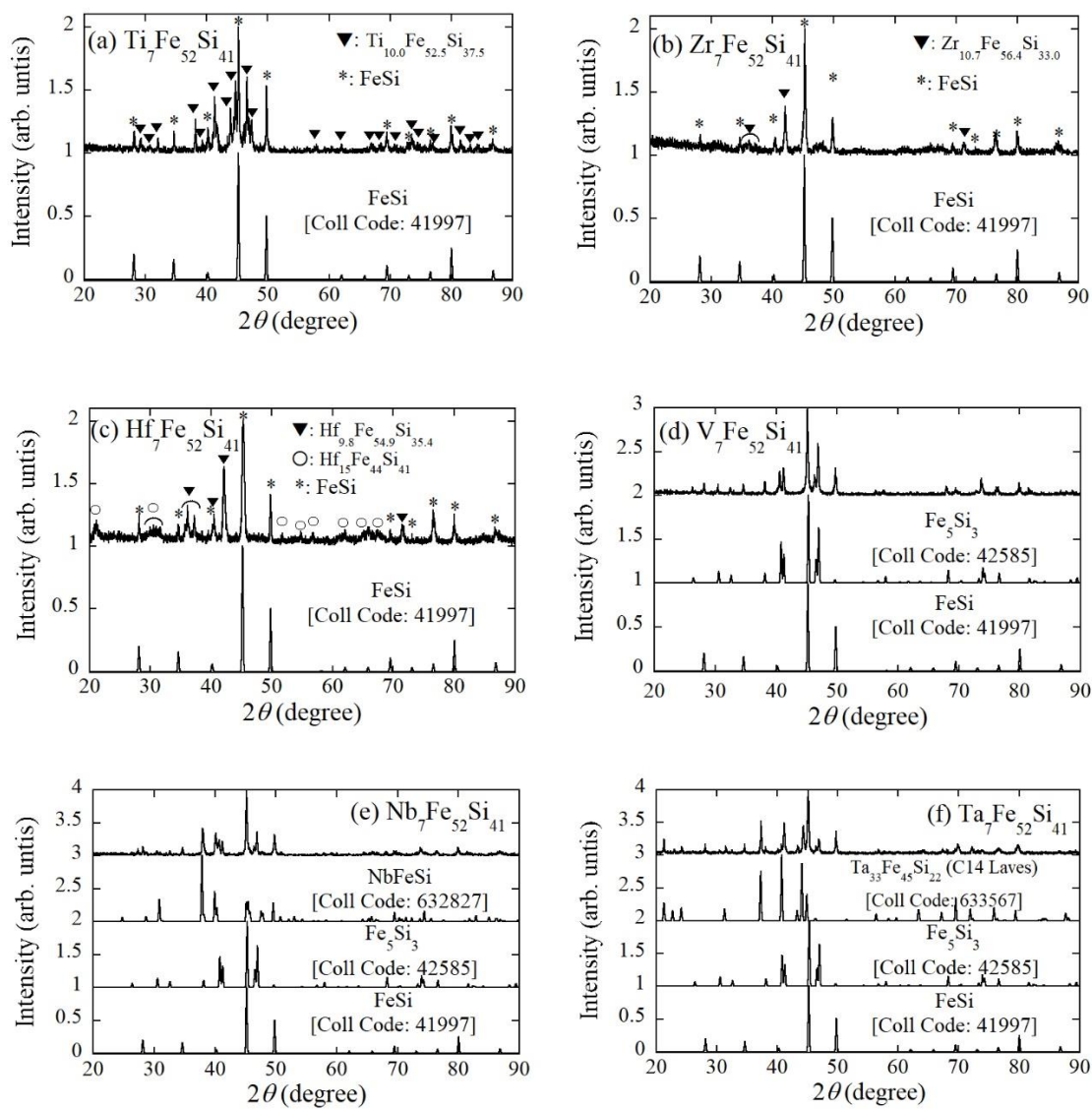

Fig.2 XRD patterns of (a) Ti$_7$Fe$_{52}$Si$_{41}$, (b) Zr$_7$Fe$_{52}$Si$_{41}$, (c) Hf$_7$Fe$_{52}$Si$_{41}$, (d) V$_7$Fe$_{52}$Si$_{41}$, (e) Nb$_7$Fe$_{52}$Si$_{41}$, and (f) Ta$_7$Fe$_{52}$Si$_{41}$, respectively. In each figure, the simulation pattern of the assigned phase with a known crystal structure is added. The origin of each pattern is shifted by an integer value for clarity.

Figs. 3(a) to 3(f) summarize the magnetic property data of TM$_7$Fe$_{52}$Si$_{41}$ (TM=Ti, Zr, and Hf). The temperature dependence of dc magnetization $M$ was measured under the external field $H$ of 100 Oe. Ti$_7$Fe$_{52}$Si$_{41}$ exhibits smaller $M$ compared to Zr$_7$Fe$_{52}$Si$_{41}$ and Hf$_7$Fe$_{52}$Si$_{41}$ (Figs. 3(a), 3(c), and 3(e)). *M-H* (*M*: magnetization) curves of Ti$_7$Fe$_{52}$Si$_{41}$ (Figs. 3(b)) showing almost linear behaviors indicate a paramagnetic character at least down to 50 K. Zr$_7$Fe$_{52}$Si$_{41}$ and Hf$_7$Fe$_{52}$Si$_{41}$ enter into a ferromagnetic state below approximately 500 K (Figs. 3(c) and 3(e)). $T_C$ (Curie temperature) values of Zr$_7$Fe$_{52}$Si$_{41}$ and Hf$_7$Fe$_{52}$Si$_{41}$ are 495 K and 488 K, respectively, which are determined by the minimum point of temperature derivative of $M$ (insets of Figs. 3(c) and 3(e)). This analysis method is commonly employed for transition metal-based ferromagnets [2,3]. $T_C$ of Zr$_7$Fe$_{52}$Si$_{41}$ is close to the reported value (533 K) [1]. Figs. 3(d) and 3(f) suggest large coercive fields $H_c$ for Zr$_7$Fe$_{52}$Si$_{41}$ and Hf$_7$Fe$_{52}$Si$_{41}$. In Zr$_7$Fe$_{52}$Si$_{41}$, $H_c$=5.3 kOe at 300 K is comparable with the reported value (4.2 kOe) [1] and is unchanged down to 200 K. At 100 K and 50 K, we confirmed the reduced $H_c$. Hf$_7$Fe$_{52}$Si$_{41}$ also shows a relatively large $H_c$ of 4.7 kOe at 300 K; however, $H_c$ gradually decreases as the temperature is lowered below 300 K. The magnetic property data of TM$_7$Fe$_{52}$Si$_{41}$ (TM=V, Nb, and Ta) are presented in Figs. 4(a) to 4(f). Each sample shows a ferromagnetic transition, as shown in Fig. 4(a), 4(c), or 4(e). The insets of Figs. 4(a), 4(c), and 4(e) showing the temperature derivative of $M$ indicate that $T_C$ values of V$_7$Fe$_{52}$Si$_{41}$, Nb$_7$Fe$_{52}$Si$_{41}$, and Ta$_7$Fe$_{52}$Si$_{41}$ are 202 K, 318 K, and 318 K, respectively. The *M-H* curves exhibit typical soft ferromagnetic character below approximately $T_C$ (Figs. 4(b), 4(d), and 4(f)). The raw data of Figs. 3(a) to 3(f) and Figs. 4(a) to 4(f) are stored in the Mendeley Data repository.

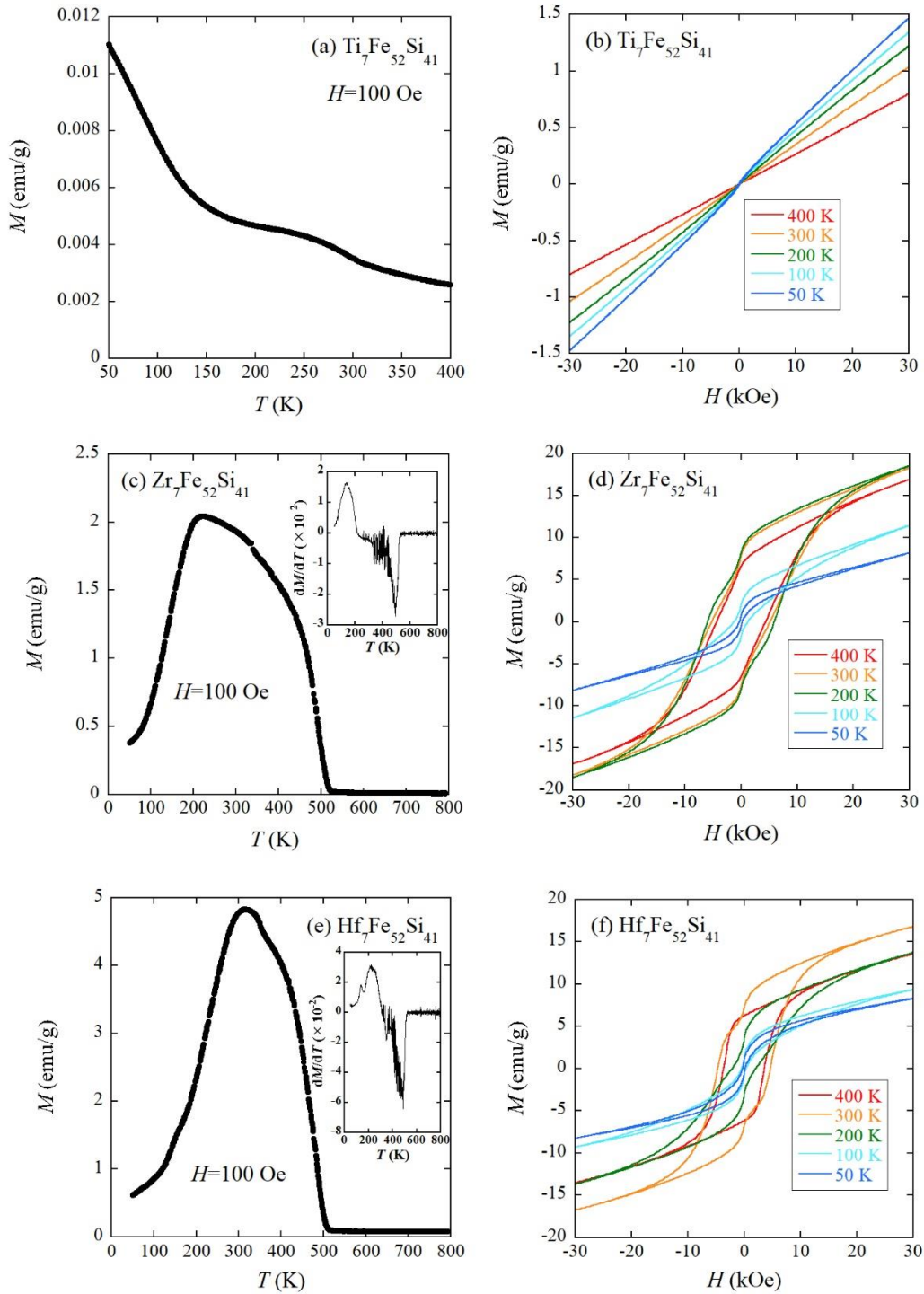

Fig.3 (a) Temperature dependence of $M$ and (b) isothermal $M$-$H$ curves at 50 K, 100 K, 200 K, 300 K, and 400 K for $Ti_7Fe_{52}Si_{41}$. The same datasets are displayed for $Zr_7Fe_{52}Si_{41}$ ((c) and (d)) and $Hf_7Fe_{52}Si_{41}$ ((e) and (f)). Insets of (c) and (e) are the temperature derivative of $M$ for $Zr_7Fe_{52}Si_{41}$ and $Hf_7Fe_{52}Si_{41}$, respectively.

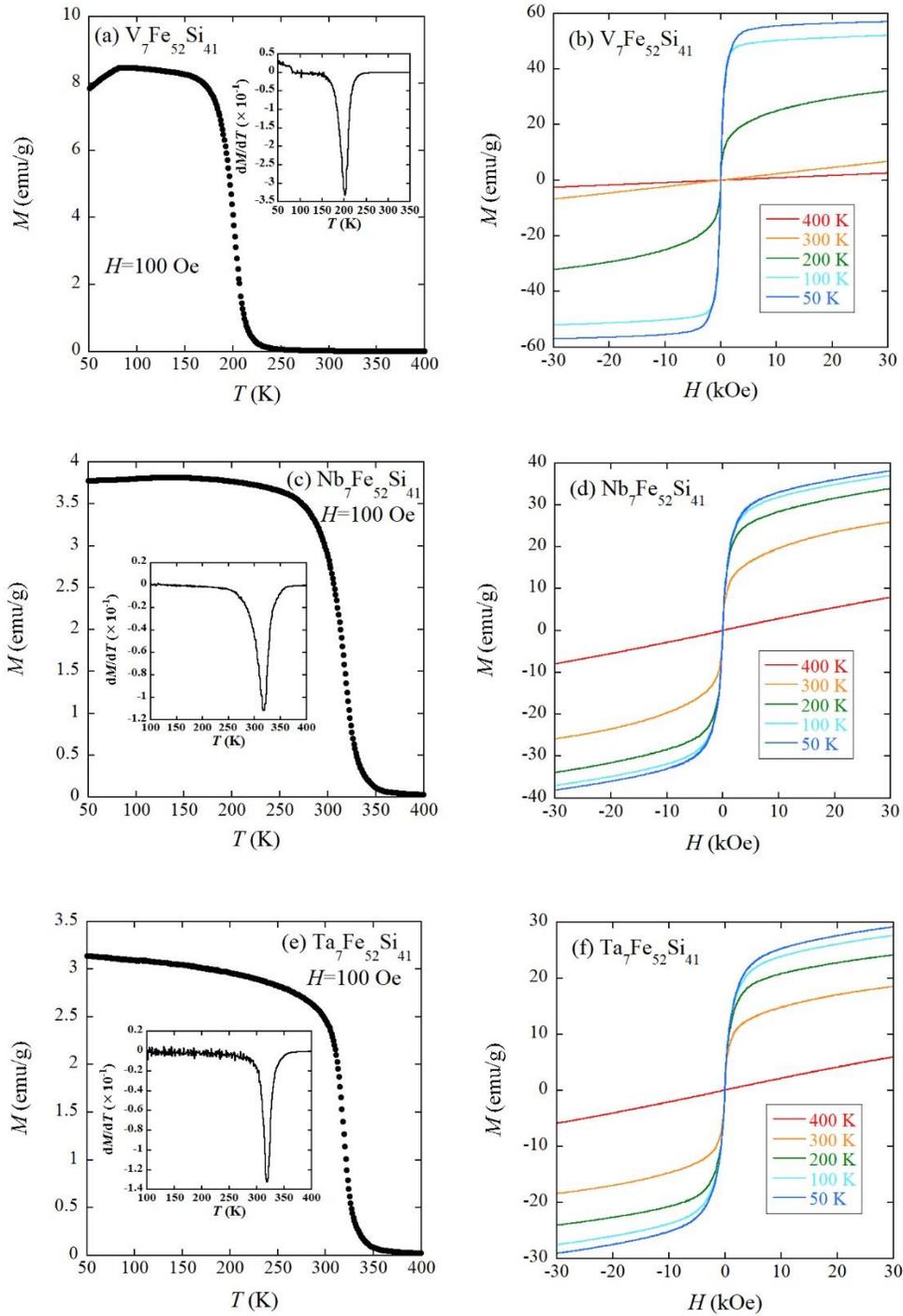

Fig.4 (a) Temperature dependence of $M$ and (b) isothermal $M$-$H$ curves at 50 K, 100 K, 200 K, 300 K, and 400 K for $V_7Fe_{52}Si_{41}$. The same datasets are displayed for $Nb_7Fe_{52}Si_{41}$ ((c) and (d)) and $Ta_7Fe_{52}Si_{41}$ ((e) and (f)). Insets of (a), (c), and (e) are the temperature derivative of $M$ for $V_7Fe_{52}Si_{41}$, $Nb_7Fe_{52}Si_{41}$, and $Ta_7Fe_{52}Si_{41}$, respectively.

**Experimental design, materials and methods**

Polycrystalline samples with a mass of 1.5 g were synthesized by a home-made arc furnace using constituent elements Ti (99.9 %), Zr (99.5 %), Hf (98 %), V (99.9 %), Nb (99.9 %), Ta (99.9 %), Fe (99.9 %), and Si (99.999 %) under Ar atmosphere. The atomic ratio was TM (=Ti, Zr, Hf, V, Nb, or Ta) : Fe : Si =7 : 52 : 41. The button-shaped samples were remelted several times on a water-cooled Cu hearth and flipped each time to ensure homogeneity. Each as-cast sample was subsequently placed in an evacuated quartz tube and annealed at 1050 ℃ for one day, followed by an air-cooling in an electric furnace. Room temperature X-ray diffraction (XRD) patterns of annealed samples were collected using an X-ray diffractometer (XRD-7000L, Shimadzu) with Cu-Kα radiation in Bragg-Brentano geometry. The phase analysis was carried out using a field emission scanning electron microscope (FE-SEM, JSM-7100F, JEOL). After checking the SEM images, the chemical composition in each area (~ 5μm × 5μm) was measured by an energy dispersive X-ray spectrometer (EDS) equipped with the FE-SEM. The chemical composition was determined by averaging several data collection points. The temperature dependence of dc magnetization $M(T)$ between 50 and 400 K was measured using a vibrating sample magnetometer (VSM) option in VersaLab (Quantum Design). The high-temperature $M(T)$ from 400 K to 800 K was measured by another VSM (TM-VSM33483-HGC, Tamakawa). In each measurement, the external field ($H$) was 100 Oe. The isothermal magnetization ($M$) curves between $H$=-30 kOe and $H$=30 kOe at 50 K, 100 K, 200 K, 300 K, and 400 K were measured using the VSM option of VersaLab.

**Ethics statements**

The authors followed universally expected standards for ethical behavior in conducting and publishing scientific research.

**CRediT author statement**

**Ryusei Ishibashi:** Fromal analysis, Investigation. **Jiro Kitagawa:** Conceptualization, Methodology, Formal analysis, Writing- Original draft preparation, Writing- Reviewing and Editing, Supervision.

**Acknowledgments**

J.K. is grateful for the support provided by Comprehensive Research Organization of Fukuoka Institute of Technology.

**Declaration of interests**

The authors declare that they have no known competing financial interests or personal relationships that could have appeared to influence the work reported in this paper.